\documentstyle[12pt,psfig,a4]{article}
\begin{document}
\def\doublespaced{\baselineskip=\normalbaselineskip\multiply\baselineskip 
  by 150\divide\baselineskip by 100}
\doublespaced
\def\sm{\mbox{$SU(3)_C\times SU(2)_L \times U(1)_Y$}\,}
\def\suu{$SU(2)_L \times U(1)_Y$\,}
\def\su{$SU(2)_l \times SU(2)_h\times U(1)_Y$\,}
\def\slash{\not{}{\mskip-3.mu}}
\def\ra{\rightarrow}
\def\lra{\leftrightarrow}
\def\bea{\begin{eqnarray}}
\def\ena{\end{eqnarray}}
\def\beq{\begin{equation}}
\def\enq{\end{equation}}
\def\cs{\cos{\theta}}
\def\sn{\sin{\theta}}
\def\tw{\tan\theta}
\def\css{\cos^2{\theta}\,}
\def\sns{\sin^2{\theta}\,}
\def\tns{\tan^2{\theta}\,}
\def\eg{{\it e.g.},\,\,}
\def\ie{{\it i.e.},\,\,}
\def\etc{{\it etc}}
\def\MWs{M^2_W}
\def\MZs{M^2_Z}
\def\MHs{m^2_H}
\def\mf{m_f}
\def\mb{m_b}
\def\mt{m_t}
\def\snshel{\sin^2{\hat{\theta}}\,}
\def\csshel{\cos^2{\hat{\theta}}\,}
\def\snsms{\overline{s}^2_{\theta}\,}
\def\cssms{\overline{c}^2_{\theta}\,}
\def\kln{\kappa_{L}^{NC}}
\def\krn{\kappa_{R}^{NC}}
\def\klc{\kappa_{L}^{CC}}
\def\krc{\kappa_{R}^{CC}}
\def\ttz{{\mbox {\,$t$-${t}$-$Z$}\,}}
\def\bbz{{\mbox {\,$b$-${b}$-$Z$}\,}}
\def\tta{{\mbox {\,$t$-${t}$-$A$}\,}}
\def\bba{{\mbox {\,$b$-${b}$-$A$}\,}}
\def\tbw{{\mbox {\,$t$-${b}$-$W$}\,}}
\def\tltlz{{\mbox {\,$t_L$-$\overline{t_L}$-$Z$}\,}}
\def\blblz{{\mbox {\,$b_L$-$\overline{b_L}$-$Z$}\,}}
\def\brbrz{{\mbox {\,$b_R$-$\overline{b_R}$-$Z$}\,}}
\def\tlblw{{\mbox {\,$t_L$-$\overline{b_L}$-$W$}\,}}
\def\ppbar{ \bar{{\rm p}} {\rm p} }
\def\pp{ {\rm p} {\rm p} }
\def\ipb{ {\rm pb}^{-1} }
\def\ifb{ {\rm fb}^{-1} }
\def\stds{\strut\displaystyle}
\def\SST{\scriptscriptstyle}
\def\TT{\textstyle}
\def\D0{D\O~}
\def\sqrts{\sqrt{s}}
\def\lsim{~{\rlap{\lower 3.5pt\hbox{$\mathchar\sim$}}\raise 1pt\hbox{$<$}}\,}
\def\gsim{~{\rlap{\lower 3.5pt\hbox{$\mathchar\sim$}}\raise 1pt\hbox{$>$}}\,}
\def\thisday{~\today ~and~ hep-ph/9810353~~}


\begin{titlepage}
\vspace{0.5cm}
\begin{flushright}
\hfill KEK-TH-596 \\
October 1998 \hfill MSUHEP-81010 
\end{flushright}
\vspace{0.5cm}
\begin{center}
\large
{Low Energy Data and a Model of Flavor Mixing}
\end{center}  
\begin{center}
{\bf Ehab Malkawi$^{a,b,}$\footnote{Currently, on leave at KEK, Japan, 
e-mail:malkawi@theory.kek.jp} and C.--P. Yuan$^c$}
\end{center}
\begin{center}
{$^a$Department of Physics,
Jordan University of Science \& Technology\\
 Irbid 22110, Jordan}
\end{center}
\begin{center}
$^b$Theory Group, KEK, Tsukuba, Ibaraki 305-0801, Japan
\end{center}
\begin{center}
{$^c$Department of Physics and Astronomy,
Michigan State University \\
East Lansing, MI 48824, USA}
\end{center} 
\vspace{0.4cm}
\raggedbottom 
\relax
\begin{abstract}
\noindent
We consider a model in which the third family fermions are subjected to 
an $SU(2)$ interaction different from the first two family fermions. 
Constrained by the $Z$-pole data, the heavy gauge boson mass is bounded 
from below to be about $1.7$ TeV at the $2\sigma$ level. In this model, 
the flavor mixing between $\tau$ and $\mu$ can be so large that 
${\rm{Br}}(\tau\rightarrow \mu \overline{\nu_\mu} \nu_\tau)/
{\rm{Br}}(\tau\rightarrow e \overline{\nu_e} \nu_\tau)$ 
and ${\rm{Br}}(\tau\rightarrow \mu \mu \mu)$ 
provide a better constraint than the LEP/SLC data in some region of
parameter space. Furthermore, flavor-changing neutral currents are
unavoidable in the quark sector of the model. Significant effects to the 
$B^0$-$\overline{B^0}$ mixing and the rare decays of the $K$ and $B$ 
mesons, such as $K^\pm \to \pi^\pm \nu {\overline \nu}$, 
$b \to s \nu {\overline \nu}$, $B_s \to \tau^+\tau^-$, $\mu^+\mu^-$ and
$B_{s,d} \to \mu^\pm \tau^\mp$, are expected.
\end{abstract} 

\vspace{0.5cm} \noindent PACS numbers:12.15.Ji, 12.60.-i, 12.60.Cn, 13.20.-
v, 13.35.-r 
\vspace{1.0cm} 
\end{titlepage} 
\newpage 

\setcounter{page}{1}
\pagenumbering{arabic}
\pagestyle{plain}

\section{The Model}

Despite the great success of the Standard Model (SM),
it offers no explanation 
of the origin of fermion flavor, or of the existence of three families.
The flavor physics of the third family is particularly mysterious for
the smallness of the mixing angles and the huge hierarchy in masses.
In this Letter, we consider a model  \cite{ehab5} 
 based on the gauge symmetry G=~$SU(3)_C\times SU(2)_l\times
SU(2)_h\times U(1)_Y$\thinspace. The third generation of fermions (top
quark, $t$, bottom quark, $b$, tau lepton, $\tau $, and
 its neutrino, $\nu_\tau $) experiences a new gauge interaction, 
instead of the usual weak interaction advocated by the SM. On the contrary,
the first and second
generations only feel the weak interaction supposedly equivalent to the SM
case. The new gauge dynamics is attributed to the $SU(2)_h$ symmetry
under which the left-handed fermions of the third generation transform in
the fundamental representation (doublets), while they remain to be singlets
under the $SU(2)_l$ symmetry. On the other hand, the left-handed fermions of
the first and second generation transform as doublets under the $SU(2)_l$
group and singlets under the $SU(2)_h$ group. The $U(1)_Y$ group is the SM
hypercharge group. The right-handed fermions only transform under the 
$U(1)_Y $ group as assigned by the SM. Finally the QCD interactions and the
color symmetry $SU(3)_C$ are the same as that in the SM.

The spontaneous symmetry-breaking of the group 
$SU(2)_l \times SU(2)_h\times U(1)_Y$\, is accomplished by introducing two
scalar matrix fields $\Sigma$ and $\Phi$, which are assigned as 
$\Sigma\sim (2,2)_0 $ and $ \Phi \sim (2,1)_1\,$.
(The scalar doublet $\Phi$ carries equivalent quantum numbers as the SM 
Higgs doublet.)
We assume that the
first stage of symmetry breaking is accomplished through the $\Sigma $
field by acquiring a vacuum expectation value $u$, i.e.,  
$
\left\langle \Sigma \right\rangle =\pmatrix{u & 0 \cr 0 & u \cr}\,.
$
The second stage is 
through the scalar $\Phi $ field by
acquiring a vacuum expectation value $v$, so 
$
\left\langle \Phi \right\rangle=\pmatrix{0 \cr v\cr}\,
$,
where $v$ is at the same order as the SM symmetry-breaking scale. Because 
of this
pattern of symmetry breaking, the gauge couplings are related to the 
$U(1)_{{\rm {em}}}$ gauge coupling $e$ by the relation 
$
{1/e^2}={1/g_l^2}+{1/g_h^2}+{1/{g^{\prime }}^2}\,\,.
$
We can define 
\begin{equation}
g_l=\frac e{\sin \theta \cos \phi }\,,\hspace{1cm} 
g_h=\frac e{\sin \theta \sin \phi }\,,\hspace{1cm}
g^{\prime }=\frac e{\cos \theta }\,,
\end{equation}
where $\theta $ is the usual weak mixing angle and $\phi $ is a new
parameter of the model. 

To derive the mass eigenstates and physical masses of the gauge bosons, we
need to diagonalize their mass matrices. 
For $g_h>g_l$ (equivalently $\tan \phi <1$), we require 
$g_h^2\leq 4\pi $ (which implies $\sin ^2\phi \geq g^2/(4\pi )\sim 1/30$) so
that the perturbation theory is valid. Similarly, for $g_h<g_l$, we require 
$\sin ^2\phi \leq 0.96$. For simplicity, we focus on the region where 
$x(\equiv u^2/v^2)$ is much larger than 1, and
ignore the corrections which are suppressed by higher powers of $1/x$. 
To the order $1/x$, the light gauge boson masses are found to be 
$
M^2_{W^\pm }=M_Z^2 {\cos ^2{\theta }\,}=
M_0^2(1- {\sin ^4\phi / x} )\,,
$
where $M_0 \equiv ev/2\sin \theta $.
While for the heavy gauge bosons, one finds 
\begin{equation}
M^2_{{W^\prime}^\pm} = M^2_{Z^\prime }=M_0^2\left( \frac x{\sin^2\phi
\cos^2\phi }+\frac{\sin^2\phi }{\cos^2\phi }\right) \,.
\end{equation}
Up to this order, the heavy gauge bosons are degenerate in mass 
because they do not mix with the hypercharge gauge boson field.

The first and second generation fermions
 acquire their masses through the Yukawa interactions to the 
$\Phi $ doublet field, while 
the third generation fermion masses have to be generated through
higher dimension operators.
Given the fermion mass matrices, one can derive their physical
masses by diagonalizing the mass matrices using bilinear unitary
transformations. For example,
In terms of the fermion mass eigenstates, we have the interaction terms:
\begin{eqnarray*}
\frac e{\sqrt{2}\sin \theta }\left( 
\begin{array}{ccc}
\overline{u_L} & \ \overline{c_L} & \overline{t_L}
\end{array}
\right) \gamma ^\mu \left[ (1-\frac{\sin ^4\phi }x)L_u^{\dagger }L_d+
\frac{\sin ^2\phi }xL_u^{\dagger }GL_d\right] \left( 
\begin{array}{c}
d_L \\ 
s_L \\ 
b_L
\end{array}
\right) W_\mu ^{+}\,+h.c.,
\end{eqnarray*}
\begin{equation}
\frac e{2\sin \theta \cos \theta }\left( 
\begin{array}{ccc}
\overline{d_L} & \ \overline{s_L} & \overline{b_L}
\end{array}
\right) \gamma ^\mu \left[ -1+\frac 23\sin ^2\theta +\frac{\sin ^4\phi }x-
\frac{\sin ^2\phi }xL_d^{\dagger }GL_d\right] \left( 
\begin{array}{c}
d_L \\ 
s_L \\ 
b_L
\end{array}
\right) Z_\mu \, ,
\label{qu2}
\end{equation}
where $L_u$ and $L_d$ are unitary matrices needed to diagonalize the up 
and down quark matrices, respectively, and
\begin{equation}
G=\pmatrix{0 & 0 & 0 \cr 0 & 0 & 0 \cr 0 & 0 & 1 \cr}\, .  \label{G}
\end{equation}

For the charged-current interactions in the quark sector, one observes that
in the case of ignoring the new physics effect, quark mixing is described by
a unitary matrix $V=L_u^{\dagger }L_d$ which is identified as the usual
Cabibbo-Kobayashi-Maskawa (CKM) mixing matrix. With the inclusion of new
physics, the mixing acquires an additional contribution proportional to 
$\sin ^2\phi /x$, with
$
{L_u^{\dagger }}GL_d={L_u^{\dagger }}L_dL_d^{\dagger }GL_d=VL_d^{\dagger
}GL_d={L_u^{\dagger }}GL_uV.
$
Therefore, we would expect the extracted values of the CKM matrix elements
to be slightly modified due to the new contributions of the model.

In this model, 
lepton mixing is an exciting possibility. However,
the almost vanishing partial decay widths of
${\mu^- \rightarrow e^- e^+ e^-} $ and 
${\mu^- \rightarrow e^- \gamma}$ have severely 
suppressed any possible mixing between the first
and second family leptons. 
Since the 
lepton number violation processes involving the third
family are not as well constrained,
and the mixing strength between leptons may be directly
related to their masses, we assume
in this paper that lepton flavor mixing
is only allowed between the second and the third family.
The strength of the lepton flavor mixing in this model is parametrized by 
$\sin \beta $. The minimal mixing corresponds to $\beta =0$,
and a maximal mixing implies $\beta =\pi/4$.
Furthermore, flavor mixing can exist in the neutrino sector 
despited that neutrinos are massless, 
which may induce an interesting effect to the neutrino 
oscillation phenomena.
As to be shown later, quark mixing must be present in the model, 
although lepton mixing is merely a possibility.
Hence, flavor-changing neutral currents
(FCNCs) are unavoidable in the quark sector.
In the following, we discuss the effect 
of the new physics predicted by this model to low-energy experiments,
as well as the constraints derived from the present data on the
parameter space of the model.

\section{Constraints From $Z$-Pole Data}

\begin{figure}[t]
\begin{center}
\leavevmode\psfig{file=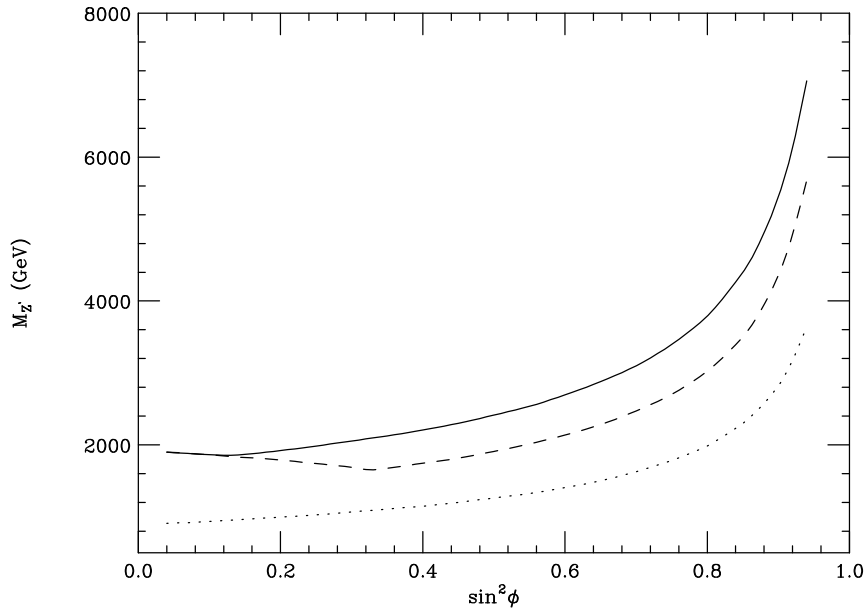,angle=90,height=8.cm}
\end{center}
\caption{The lower bound on the heavy mass $Z^\prime$ as a function of 
 $\sin^2\phi$ at the $2\sigma$ level. 
     Solid curve: including all $Z$-pole data and assuming no lepton mixing.
     Dashed curve: including all $Z$-pole data 
and assuming maximal lepton mixing ($\sin^2\beta=0.5$).
     Dotted curve: only including the hadronic 
measurements in the fit and 
     assuming no lepton mixing.}
\label{fig1}
\end{figure}

\noindent                                    
Following Ref.~\cite{ehab5}, we calculate the changes in the 
relevant physical 
observables relative to their SM values to leading order in $1/x$.
Namely, any observable $O$ can be written as  
$
O=O^{\rm{SM}}\left(1+\delta O\right ) \, ,
$      
where $O^{\rm{SM}}$ is the SM prediction including the
one-loop SM correction, and $\delta O$ 
represents the new physics effect to leading order in $1/x$.
using the most recent LEP and SLC
measurements \cite{lep}
(including the total width of the $Z$
boson $\Gamma_Z$, $R_e$, $R_\mu$, $R_\tau$, 
the vector $g_{Ve}$ and the 
axial-vector $g_{Ae}$ couplings of the electron, the ratios 
$g_{V(\mu,\tau)}/g_{Ve}$, $g_{A(\mu,\tau)}/g_{Ae}$, 
$A_{FB}^e$, $A_{FB}^\mu$, $A_{FB}^\tau$, $A_e$, $A_\tau$, $M_W$, 
the hadronic cross section $\sigma_h^0$, $R_b$, $R_c$, $A_b$, $A_c$, and 
$A_{FB}^c$), and the SM predictions of the above observables 
\cite{hagiwara},
we find that the precision $Z$-pole data 
requires the heavy gauge boson mass to be
larger than about $1.9$ TeV at the $2\sigma$ level, and
the parameter $x$ to be larger than 20 for $\beta=0$.
In Figure 1 (solid curve) we show the minimal 
$Z^\prime$ mass as a function of $\sin^2\phi$
at the $2\sigma$ level, assuming no mixing in the 
lepton sector. 
The $Z$-pole observable that imposes the most stringent constraint 
on the model is $R_\tau$, which is the ratio of the partial decay widths of 
$Z$ boson into the $\tau^+ \tau^-$ and the hadronic modes.
The measurement of $\Gamma_Z$ also plays 
an important role secondary to $R_\tau$, especially for small $\sin^2\phi$,
due to the high precision of data. 
We find that $x$ can be as small as 20 for 
the smallest allowed 
value of $\sin^2\phi$ ($=0.04$), and it increases as $\sin^2\phi$ 
increases.
Furthermore, the quantity $\sin^2\phi/x$ is constrained by data to be 
less than about $2\times 10^{-3}$ for a large range of $\sin^2\phi$.
As also shown in Figure 1 (dotted curve), 
without including the leptonic observables from the $Z$-pole data, 
the bounds on $M_{Z^\prime}$ and $x$ is about 900 GeV 
and 5, respectively.
 (In this case,
the important constraint is coming from the measurement of $R_b$.)

If a mixing between $\mu$ and $\tau$ is allowed,
we also need to include the limit on the branching ratio of
$Z\rightarrow \mu \tau$ from LEP/SLC data \cite{data}.
In Figure 1 (dashed curve), we depict the new constraints on 
$\sin^2\phi$ and $M_{Z^\prime}$ for 
$\sin^2\beta=0.5$. We find that the lower limit on the heavy mass is
 reduced to $M_{Z^\prime}\approx 1.7$ TeV.
The reason for this slightly
lower bound is due to the reduced non-universal 
effect in $R_\tau$. 
We find that for the smallest allowed value of 
$\sin^2\phi=0.04$, the value of $x$ can be as low as 20,
which is similar to the case of $\sin^2\beta=0$.
The reason is that 
for small values of $\sin^2\phi$ ($<0.2$), the measurement $\Gamma_Z$, which is 
independent of the mixing angle $\sin^2\beta$, plays 
the important role in constraining the parameter $x$. 
Again, 
the quantity $\sin^2\phi/x$ is found to be less than about $0.3\%$.

\section{Constraints From Low-Energy Data}

Even though the $Z$-pole data already impose significant constraints, 
this model has a rich structure that can be 
further examined at much lower energy scales. 
In the following sections, 
we would like to examine those constraints obtained from 
 the low-energy hadronic, leptonic, and semi-leptonic data. 
We will concentrate on the very low-energy regime, i.e., physics at
zero-momentum transfer, and examine whether 
the parameters of the model can be better constrained 
than those imposed by LEP and SLC data. 
For clarity, we shall separately discuss below the effects from the 
lepton and quark sectors to lepton number violation phenomena,
as well as to kaon and bottom physics.

\subsection{The Lepton Sector}

It is easy to see that if
the third family lepton does not mix with the 
first family lepton, then the partial decay width of
$\mu \rightarrow eee$ and $\mu \rightarrow e \gamma$
will not be modified.  
However, if the lepton mixing involves the third family,
then the lifetime of the $\tau $ lepton will be modified.
The only modification to the total decay width, to the order $1/x$, 
is coming from the partial 
decay width 
$\Gamma\left( \tau ^{-}\rightarrow \mu ^{-}
\overline{\nu _\mu }\nu _\tau\right)$. 
We find
\begin{equation}
\frac{{\rm{Br}} \left( \tau ^{-}\rightarrow 
\mu ^{-}\overline{\nu _\mu }\nu _\tau
\right)}
{{\rm{Br}}\left( \tau ^{-}\rightarrow e ^{-}\overline
{\nu _e }\nu _\tau \right)}=f(m_\mu/m_\tau)
\left( 1+\frac{3\cos ^2\beta \sin ^2\beta}{x}\right)\, , 
\end{equation}
where $f(m_\mu/m_\tau)$ is a phase factor \cite{park} given by 
$
f(y)=1-8y^2+8y^6-y^8-24y^4\ln(y) \,.
$

After comparing with data \cite{data}, 
we find that the above ratio
constrains the parameter $x>27\sim 48$.  
Also the lepton number violation
process $\tau \rightarrow \mu \mu \mu $  
provides the constraint $x>37$ consistent with the above measurement. 
Therefore, the above two measurements give a stronger constraint than the  
$Z$-pole data for $\sin^2\phi <0.1$.
On the other hand, giving its present experimental limit,
the decay process $\tau\rightarrow \mu \gamma$ has not yet played a 
significant role.
Nevertheless, if the above discussed decay widths can 
be measured to a better accuracy 
in future experiments, they can further test the proposed model.
Hereafter, when
 discussing the predictions of our model to low-energy processes, 
we will consider $x\geq 20$ for $\sin^2\beta=0.0$, and
$x\geq 48$ for $\sin^2\beta=0.5$.

\subsection{The Quark Sector}

To completely describe the interactions of 
gauge bosons and quarks, it requires two 
mixing matrices $L_u$ and $L_d$ because both the up- and down-type 
quarks are massive. 
As noted in Eq.~(\ref{qu2}) that
the neutral-current mixing matrices 
($L_u^{\dagger }GL_u$ and $L_d^{\dagger }GL_d$)
are related to the charged-current mixing matrix ($L_u^{\dagger }L_d$).
Because of the experimental evidence of the CKM matrix in charged 
currents, FCNCs must occur in this model between 
the interaction of quarks and gauge bosons,
and can be realized in three different ways: 
(i) in the down-quark sector only, (ii) in the up-quark sector 
only, and (iii) in both sectors. All these
three possibilities have to confront the existing 
low-energy data.
In the following, we investigate these three possibilities, separately.

First, we consider the case that only down-type quarks can mix, 
so that $L_u^{\dagger }GL_u=G$, and
$L_d^{\dagger }GL_d=V^{\dagger}GV$.
This implies
that FCNC processes are completely determined by the CKM
matrix $V$, in addition to $\sin^2\phi $ and $x$.  
Since the matrix elements of 
$V^{\dagger }GV$ are naturally small, we generally do not expect 
large effects in FCNC processes.
Under the assumption that the $\tau$ lepton does not mix with the 
electron $e$,
the decay width of $K^+ \rightarrow \pi^0 e^+ \nu_e$ is not modified
at tree level.
However, $K^+ \rightarrow \pi^+ \nu \overline{\nu}$ 
will be modified through the FCNC interaction
$s\rightarrow d\, Z \rightarrow d \nu \overline{\nu}$. 
The expected branching ratio, normalized to 
the predicted branching ratio for $K^+ \rightarrow \pi^0 e^+ \nu_e $, 
can be written as  
\begin{equation}
\frac{{\rm{Br}}(K^+ \rightarrow \pi^+ \nu \overline{\nu})}
{{\rm{Br}}(K^+ \rightarrow \pi^0 e^+ \nu_e)}=
\frac{1}{4x^2} \left( \frac{|V_{td}|^2 |V_{ts}|^2}{|V_{us}|^2}\right)\, .
\end{equation}
For $x=20$, we find that 
${\rm{Br}}(K^+ \rightarrow \pi^+ \nu \overline{\nu})$
is a few times larger than the SM prediction. 
In the above result we summed over all neutrino flavors, therefore, the
$\sin^2\beta$ dependence cancels.
Furthermore, since the above observables do not depend on the parameter
$\sin^2\phi$, they can directly constrain the parameter $x$ of the model.
The effect to the $K^0$-$\overline{K^0}$ mixing is of the same order as the 
SM prediction, which can only be useful if the long distance contribution
can be better understood theoretically.

This model also predicts non-standard effects for bottom
physics. The important process to consider is 
$B^0_q$-$\overline{B^0_q}$ mixing where new effect is expected 
to occur at tree level. 
After substituting all the relevant variables by their numerical values,
we find
\begin{equation}
\frac{\Delta M_{B_q}}{(\Delta M_{B_q})^{\rm{SM}}} \simeq
\frac{72}{x}\, .
\end{equation}
Given the constraint on $x$ ($>20$) imposed by the $Z$-pole data,
without lepton flavor mixing (i.e., $\sin^2 \beta=0$),
the new contribution can be a few times of the SM predictions.
In the case that $\tau$ and $\mu$  
mix with the maximal strength (i.e., $\sin^2 \beta=0.5$ and $x>48$), 
the new contribution to $B^0_q$-$\overline{B^0_q}$ mixing 
is expected to be less than $150\%$.
Furthermore, our model predicts tree-level contribution to the process 
$b\rightarrow s \nu \overline{\nu}$, whose
branching ratio, when normalized by
${{\rm{Br}}(b \rightarrow c \mu^- \overline{\nu_\mu})}$,
is given as
\begin{equation}
\frac{{\rm{Br}}(b\rightarrow s \nu \overline{\nu})}
     {{\rm{Br}}(b \rightarrow c \mu^- \overline{\nu_\mu})}
=\frac{1}{4x^2} 
\frac{|V_{ts}V_{tb}|^2}{|V_{cb}|^2 f(z)} \, ,
\label{eq:bsnn}
\end{equation}
where $f(z)=1-8z^2+8z^6-z^8-24z^4\ln(z)$ and $z=m_c/m_b$ 
\cite{buras,ahmad}.
The typical size of the predicted branching ratio is 
listed in Table 1.
Another interesting process to consider is the decay
$B_{s,d}\rightarrow \ell^+ \ell^-$. 
At tree level, the decay rate is given by
\begin{equation}
\Gamma(B_q\rightarrow \tau^+ \tau^- )=
\frac{G_F^2 f_{B_q}^2 m_{B_q} m_\tau^2 |V_{tb}V_{tq}|^2}{4\pi x^2} 
      {\left(\cos^2\beta -4\sin^2\theta\sin^2\phi\right)}^2
       {\left(1-\frac{4m_\tau^2}{m^2_{B_q}}\right)}^{3/2}\, .
\end{equation}
Using the values $\cos^2\beta=0.5$, $\sin^2\phi=0.04$, $m_{B_s}=5.369$ GeV, 
and $f_{B_s}=0.23$ GeV \cite{ali2}, 
we find that the current data on the branching ratios of 
$B_{s,d} \rightarrow \tau^- \tau^+, \mu^- \mu^+$
does not impose a better constraint on the model 
than that by the $Z$-pole data. 
Since this model also predicts the non-SM decay modes, such as
$b \rightarrow s \mu^\pm \tau^\mp$ and 
$B_{s,d} \rightarrow \mu^\pm \tau^\mp$,
with comparable branching ratios,
they should be measured to test the model prediction 
on the lepton flavor mixing dynamics (i.e., $\sin^2\beta$ dependence).
As a summary to this scenario,
in Table 1 we tabulate the predictions of our model for 
various processes in two cases: (i) $\sin^2\beta=0.0$ and 
$x=20$, (ii) $\sin^2\beta=0.5$ and $x=48$. 
For both cases we set $\sin^2\phi=0.04$.

Under the scenario that only up-type quarks can mix, 
there will be no non-standard effect present in the 
hadronic decays of $K$, $D$ and $B$ mesons.
This is because in the pure hadronic interaction, the
new physics effect is only expected in processes that
$\Delta B$ vanishes, i.e., the number of
$b$ quark must be the same in the initial and final states
of a reaction.
It is easy to see that
new effects in the charged-current semi-leptonic decays 
are only expected in the $b$-quark system.
As an example, 
it can modify the decay branching ratio of
$B_d^0\longrightarrow D^{-}\ell ^{+}\nu$ 
and  
$B_s^0\longrightarrow D_s^{-}\ell ^{+}\nu $ 
to be
\begin{equation}
{\rm{Br}(B^0\longrightarrow D^{-}\ell ^{+}\nu)}=
{\rm{Br}}^{\rm{SM}}(B^0\longrightarrow D^{-}\ell ^{+}\nu)
\left( 1+\frac{2}{x}\right) ,
\end{equation}
where all the three lepton (including neutrino) flavors are included.
(Note that the $\sin^2\beta$ dependence cancels.)
However, with $x>20$, imposed by the $Z$-pole data, we do not expect the 
new physics effects to exceed $10\%$.
Furthermore, the unitarity condition of the CKM matrix 
will be modified, but its change is extremely 
small for the values of $x$ that
agree with $Z$-pole data.
At one-loop level, the $K^0$-$\overline{K^0}$, 
$B^0$-$\overline{B^0}$ mixing and the decay width of 
$b \rightarrow s \gamma$ can be modified, but
the one-loop effects are generally small and do 
not exceed the level of $0.4\%$ compared with the SM prediction.
 Although FCNC decay of charm meson is expected to be modified
at tree level, 
the non-standard effect is again small because of the natural suppression
imposed by the FCNC couplings (which are the products of 
CKM matrix elements).
The only suspected new effect in the $b$-quark system is through the 
$\Upsilon(1S)$ decay, which 
proceeds through 
$b\overline{b}\rightarrow \gamma, Z, Z^\prime \rightarrow \mu ^{+}\mu ^{-}$.
We find the expected maximal deviation in 
the ratio $\Gamma(\Upsilon(1S)\rightarrow \tau^+ \tau^-)/
     \Gamma(\Upsilon(1S)\rightarrow \mu^+ \mu^-)$ is less than $\pm0.1\%$
for $x\geq 20$. It vanishes for the maximal lepton mixing scenario,
i.e. for $\sin^2\beta=0.5$, and  
the partial decay width of 
$\Upsilon(1S)$ into the $\mu^\pm \tau^\mp$
has to be measured to test this scenario. 

Under the general mixing scenario,
the model requires two additional free (real) parameters,
which can be chosen to be any two of the 31, 32 and 33 elements 
(i.e., $d_{31}$,$d_{32}$ and $d_{33}$) of
the $L_d$ mixing matrix, 
 although additional phases can be introduced to 
generate a new source of CP violation.
(Note that the unitarity condition of the matrix $L_d$ requires
$|d_{31}|^2+|d_{32}|^2+|d_{33}|^2=1$.)
It turns out that depending on the values of the parameters, 
sizable effects in various FCNC processes are expected.
In Table 2, we summarize the results under this scenario 
by giving the constraints on the mixing parameters, extracted from
various experiments. 
Similar to the discussions given for the other two scenarios, the general 
mixing scenario also allows lepton number violation processes, such as 
$B_{s,d}\rightarrow \mu^\pm \tau^\mp$ and 
$b \rightarrow s \mu^\pm \tau^\mp$.
Since their branching ratios are of the same order as those for the 
$\tau^+\tau^-$ mode, they can be very useful for further testing the model.

Finally, we note that in the SM
neither $L_u$ nor $L_d$ can be separately determined, and only
the CKM matrix $V$, which is the product of 
$L_u^\dagger$ and $L_d$, can be measured experimentally.
However, in this model, the elements in the third column of 
the $L_{u,d}$ mixing matrices can be determined, and can be 
further constrained by including the other low-energy data.
In conclusion, we have shown that the current low energy data 
(including the $Z$-pole data) still allows the third family fermions to 
have different gauge interactions from the first and the second 
family fermions, though the mass of the heavy gauge bosons in this case
has to be larger than 1.7 TeV. 
Fortunately, as summarized in Tables 1 and 2, 
many of the predicted new effects of this model can
be further tested at the Tau/Charm, Kaon and Bottom factories.

\vspace{0.3cm}
\noindent
{\bf Acknowledgments}

E.M. would like to thank K. Hagiwara, Y. Okada,
for useful discussion and comments.
He also thanks the Matsumae International Foundation for 
offering the Fellowship.
Part of this work was supported by the U.S.~NSF 
under grant PHY-9802564.


\begin{table}
\caption{Predictions of various decay rates and mixing in the SM and this
  model under the $d$-quark mixing scenario.  
  Case I: $\sin^2\beta=0$, $x=20$, $\sin^2\phi=0.04$. 
  Case II: $\sin^2\beta=0.5$, $x=48$, $\sin^2\phi=0.04$.} 

\vspace{0.1in}

\begin{tabular}{|c|c|c|c|c|}  \hline\hline
Process &  Data & SM  &
\multicolumn{2}{|c|} {\mbox{d-type mixing}} \\ \hline
    --    &  --    & --  &  I  & II \\ \hline
$\frac{{\rm{Br}}(\tau^-\rightarrow \mu^- \overline{\nu_\mu} \nu_\tau)}
{{\rm{Br}}(\tau^-\rightarrow e^- \overline{\nu_e} \nu_\tau)}$
& $0.976\pm 0.006$       & 0.9729 
& 0.9729 & 0.9881 \\
${\rm{Br}}(\tau^-\rightarrow \mu^- \mu^- \mu^-)$  
&$ < 1.9 \times 10^{-6}$   & 0   
& 0   &  $1.1\times 10^{-6}$ \\
${\rm{Br}}(\tau \rightarrow \mu \gamma)$
& $<4.2\times 10^{-6}$ & 0 
& 0  & $ 1.7\times 10^{-8}$\\
${\rm{Br}}(K^0_L \rightarrow \mu^+ \mu^-)$  
& $(7.2\pm 0.5)\times 10^{-9}$ & $\sim 7\times 10^{-9}$
& $1.3\times 10^{-10}$ & $3.4\times 10^{-9}$\\  
${\rm{Br}}(K^+ \rightarrow \pi^+ \nu \overline{\nu})$  
& $4.2 ^{+9.7}_{-3.5}\times 10^{-10}$ & $(9.1 \pm 3.8)\times 10^{-11}$ &
$2.8 \times 10^{-10}$ & $4.8 \times 10^{-11}$ \\
$\Delta M_K \, (ns^{-1})$ 
&$5.311\pm 0.019$ & $2.23 \sim 7.43$ & 
$2.6\sim 8.9$ & $2.4 \sim 8.0$\\
%
%
$\Delta M_{B_s} \, (ps^{-1})$       
& $>10.2 $ & $ 1\sim 15$  &  
$5 \sim 69$  & $3 \sim 37$ \\
${\rm{Br}}(b \rightarrow s \mu^+\mu^-)$ 
& $<5.8 \times 10^{-5}$ & $\sim 7\times 10^{-6}$ & 
$1.6\times 10^{-7}$ & $9.2 \times 10^{-6}$ \\
${\rm{Br}}(b \rightarrow s \nu\overline{\nu})$ 
& $<3.9 \times 10^{-4}$ & $\sim 4.2\times 10^{-5}$ & 
$2.3\times 10^{-4}$ & $4.0 \times 10^{-5}$ \\
${\rm{Br}}(b \rightarrow s \mu^\pm \tau^\mp)$ 
& ? &  0 
& 0 & $2.0 \times 10^{-5}$ \\
${\rm{Br}}(B_d \rightarrow \mu^+\mu^-)$ 
& $<8.6 \times 10^{-7}$ &  $ 2.1\times 10^{-10}$ & 
$3.2\times 10^{-11}$ & $8.8 \times 10^{-10}$ \\
${\rm{Br}}(B_s \rightarrow  \mu^+\mu^-)$ 
& $<2.6 \times 10^{-6}$ & $ 4.3 \times 10^{-9}$ & 
$6.1\times 10^{-10}$  & $1.7 \times 10^{-8}$ \\
${\rm{Br}}(B_d \rightarrow  \mu^\pm \tau^\mp)$ 
& $<8.3 \times 10^{-4}$ & 0 & 
0  & $4.0 \times 10^{-7}$ \\
${\rm{Br}}(B_s \rightarrow  \mu^\pm \tau^\mp)$ 
& ? & 0 & 
0  & $7.7 \times 10^{-6}$ \\
${\rm{Br}}(B_d \rightarrow \tau^+\tau^-)$ 
& ? &  $ 4.3\times 10^{-8}$ & 
$2.6\times 10^{-6}$ & $1.0 \times 10^{-7}$ \\
${\rm{Br}}(B_s \rightarrow  \tau^+\tau^-)$ 
& ? & $ 9.1 \times 10^{-7}$ & 
$5.0\times 10^{-5}$  & $2.0 \times 10^{-6}$ \\
$\Upsilon(1S) \to \mu^\pm \tau^\mp $ 
&? & 0 &
0 & $ 4\times 10^{-10}$ \\       
\hline
\end{tabular}
\end{table}

\begin{table}
\caption{Constraints on the quark mixing parameters under the general mixing 
scenario.}

\vspace{0.1in}

\begin{tabular}{|c|c|c|c|}  \hline\hline
Process   &   $\sin^2\beta$  & $\sin^2\phi$ & Constraint  \\ \hline
${\rm{Br}}(K^+\rightarrow \pi^+ \nu\overline{\nu})$ 
& independent & independent & 
$|d_{31}d_{32}|/x \lsim 1.0\times 10^{-4}$ \\
${\rm{Br}}(b\rightarrow s \nu\overline{\nu})$ 
& independent & independent &
$|d_{32}d_{33}|/{x}\lsim 2.9\times 10^{-3}$ \\
${\rm{Br}}(b\rightarrow s \mu^+\mu^-)$ 
& 0.5 & 0.04 &
$|d_{32}d_{33}|/{x}\lsim 2.3\times 10^{-3}$ \\
${\rm{Br}}(B_d\rightarrow \mu^+\mu^-)$ 
& 0.5& 0.04 &
$|d_{31}d_{33}|/x \lsim 9.1 \times 10^{-3}$ \\
${\rm{Br}}(B_s\rightarrow \mu^+\mu^-)$ 
& 0.5 & 0.04 &
$|d_{32}d_{33}|/{x}\lsim 1.2 \times 10^{-2}$ \\ \hline
\end{tabular}
\end{table}


\end{document}